\documentclass[aps,prd,twocolumn,showpacs,nofootinbib]{revtex4}
\usepackage{latexsym}
\usepackage{amsmath,amsfonts}
\usepackage{amsbsy}
\usepackage{mathrsfs}
\usepackage{color}

\usepackage{psfrag}

\usepackage{enumerate}

\usepackage{amsmath,amssymb,calc,amsfonts}
\usepackage{latexsym}

\newcounter{mnotecount}[section]

\usepackage{graphicx,calc,epsfig}
\def\ut#1{\rlap{\lower1ex\hbox{$\sim$}}#1{}}

\newcommand{\be}{\nopagebreak[3]\begin{equation}}
\newcommand{\ee}{\end{equation}}
\newcommand{\ba}{\nopagebreak[3]\begin{eqnarray}}
\newcommand{\ea}{\end{eqnarray}}
\newcommand{\norma}[1]{\ensuremath \left\lVert\, #1 \,\right\rVert}
\DeclareFontFamily{U}{rsfs}{}         
\DeclareFontShape{U}{rsfs}{m}{n}{<5> rsfs5 <6><7> rsfs7          %
  <8><9><10><10.95><12><14.4><17.28><20.74><24.88> rsfs10}{}     %
\DeclareMathAlphabet{\mathfs}{U}{rsfs}{m}{n}                     %
\newcommand{\mfs}[1]{\mathfs {#1}}                               %
\newcommand{\n}{{\nonumber}}
\newcommand{\va}{\scriptscriptstyle}

\newcommand{\sH}{{\mfs H}}
\newcommand{\sL}{{\mfs L}}

\newcommand{\sO}{{\mfs O}}

\def\i{i}

\def\pb#1{\rlap{\lower1.5ex\hbox{$\longleftarrow$}}{#1}}
\def\dpb#1{\rlap{\lower1.5ex\hbox{$\Longleftarrow$}}{#1}}
\def\spb#1{\rlap{\lower1.5ex\hbox{$\leftarrow$}}{#1}}
\def\sdpb#1{\rlap{\lower1.5ex\hbox{$\Leftarrow$}}{#1}}

\definecolor{blue}{rgb}{0,0,1}
\definecolor{green}{rgb}{0,1,0}
\definecolor{red}{rgb}{1,0,0}
\definecolor{vio}{rgb}{1,0,1}
\definecolor{ama}{rgb}{1,1,0}

\begin{document}



\title{A local first law for black hole thermodynamics}

\date{\today}
\author{Ernesto Frodden$^{1,3}$}
\author{Amit Ghosh$^2$}
\author{Alejandro Perez$^3$}

\affiliation{$^1$Departamento de F\'{\i}sica, P. Universidad Cat\'olica de Chile, Casilla 306, Santiago 22, Chile.
$^2$Saha Institute of Nuclear Physics, 1/AF Bidhan Nagar,
700064 Kolkata, India.\\
$^3$ Centre de Physique Th\'eorique, Aix-Marseille Univ, CNRS UMR 6207, Univ Sud Toulon Var,\\ 13288 Marseille Cedex~9, France.\\
}


\begin{abstract}
%
We first show that stationary black holes satisfy an extremely simple local form of the first law \[\delta E=\frac{\overline\kappa}{8 \pi} \delta A\] where the thermodynamical energy $E=A/(8\pi \ell)$ and (local) surface gravity $\overline \kappa=1/\ell$, where $A$ is the horizon area and $\ell$ is a proper length characterizing the distance to the horizon of a preferred family of local observers suitable for thermodynamical considerations. Our construction is extended to the more general framework of isolated horizons. The local surface gravity is universal.  
This has important implications for semiclassical considerations of black hole physics as well as for the fundamental quantum description arising in the context of loop quantum gravity.
\end{abstract}


\maketitle


Hawking's semiclassical calculations
\cite{Hawking:1974sw} imply that large black holes (BH) produced by gravitational collapse behave like perfect black bodies at Hawking temperature $T_{\va H}$ proportional to their surface gravity once they have reached their stationary equilibrium state. Moreover different neighbouring stationary states are related by the first law of BH mechanics from which black holes
can be assigned an entropy $S=A/4\ell_p^2$ where $\ell_p=\hbar^{1/2}$ (in units $G=c=1$) is the Planck length and $A$ is the classical area of the event horizon.

A complete statistical mechanical account of the thermal properties of BHs from
 quantum degrees of freedom remains an important challenge
for all candidate theories of quantum gravity. Statistical entropy has been calculated
in string theory \cite{strings} and loop quantum gravity \cite{Corichi:2009wn}, yet in both cases significant
gaps remains to be filled.

An important difficulty in dealing with black holes in quantum gravity is that, as they evaporate, the usual
definition based on global structure of space-time is ill-posed. This has been recently clearly
 illustrated in the context of two-dimensional models \cite{Ashtekar:2008jd}. Nevertheless, one would expect that
the physical notion of a large black hole radiating very little and, thus, remaining
close to equilibrium for a long time could be characterized in a suitable way and that such a characterization should help in studying the appropriate semiclassical regime of the underlying quantum theory.

Such quasilocal characterization of black holes is provided by  {\em isolated horizon}s \cite{Ashtekar:2004cn}. Isolated horizons (IH) capture
the main local features of BH event horizons while being of a quasilocal nature itself. In particular, isolated horizons satisfy a quasilocal version of the first law \cite{aa}
\be
\delta E_{\va IH}=\frac{\kappa_{\va IH}}{8\pi}\delta A+\Omega_{\va IH} \delta J_{\va IH}+\Phi_{\va IH} \delta Q_{\va IH}
\ee
where $E_{\va IH}$, $J_{\va IH}$ and $Q_{\va IH}$ are suitable quasilocal energy, angular momentum, and charge functions, while $\kappa_{\va IH},  \Omega_{\va IH}$, and $\Phi_{\va IH}$ are local notions of IH surface gravity, angular velocity and electrostatic potential. 
The previous equation comes from the requirement that time evolution which respects the IH boundary conditions be Hamiltonian\cite{aa}. 
The first law implies that the IH energy is $E_{\va IH}$ must be function $E(A, J_{\va IH}, Q_{\va IH})$. The integrability conditions for $E_{\va IH}$ stemming from the previous 
phase space identity imposes the usual restrictions on the `intensive' quantities.
Beyond these conditions the first law of IH does not give a preferred notion of energy of the horizon: this is a limitation 
for  statistical mechanical descriptions of quantum BHs. 

In this letter we show that the above indeterminacy disappears if one fully develops the local perspective
from which IH were defined in the first place. In fact stationary BHs (and more generally IHs) satisfy the local 
first law
\begin{eqnarray}\label{tutiri}
\delta E =\frac{\overline\kappa}{8\pi}\delta A,
\end{eqnarray}
where 
$E_{\va IH}=E$, $\overline \kappa=\ell^{-1}$  with  $\ell^2\ll{A}$ a proper length intrinsic to our analysis. The previous
equation can be integrated, thus providing a notion of horizon energy $E=\frac{A}{8\pi\ell}$ which is precisely the one 
to be used in statistical mechanical considerations by local observers.  We first show the  validity of (\ref{tutiri})  for stationary black holes and later extend the 
proof for IHs. The area as a notion of energy has been
evoked on several occasions in the context of BH models in loop quantum gravity \cite{Krasnov:1997yt, Ghosh:2011fc}. The results of this paper put these analysis on firm ground.

\section{A Local First Law}


We first  study the thermodynamic properties of  Kerr-Newman BHs
 as seen by stationary observers $\mfs O$, located right outside the horizon at a small
proper distance $\ell\ll r_+$. They follow integral
curves of the Killing vector field
\begin{eqnarray}
\chi=\xi+\Omega\,\psi=\partial_t+\Omega\,\partial_\phi,
\end{eqnarray}
where $\xi$  and $\psi$ are the Killing fields associated with the stationarity and axisymmetry of Kerr-Newman spacetime respectively, while $\Omega$ is the horizon angular velocity 
\be \Omega=\frac{a}{r^2_++a^2}.\ee
The four-velocity of $\mfs O$ is given by
\begin{align}\label{keyy}
&u^a=\frac{\chi^a}{\|\chi\|}.
\end{align}
These observers are the unique stationary ones that coincide with the {\em locally non-rotating observers} of \cite{Wald:1984rg} or ZAMOs of \cite{Thorne:1986iy} as $\ell\to 0$. As a result, the angular momentum of these observers is not exactly zero, but $o(\ell)$. 
Thus they are at rest with respect to the horizon which makes them the preferred observers for studying thermodynamical issues from 
a local perspective.

Standard arguments lead to the so-called first law of BH mechanics
that relates different nearby stationary BH spacetimes of
Einstein-Maxwell theory
\begin{eqnarray}\label{1st}
\delta M=\frac{\kappa}{8\pi} \delta A +\Omega \delta J+\Phi  \delta Q,
\end{eqnarray}
where $M$, $J$ and $Q$ are respectively the total mass, angular momentum and charge of the spacetime, $A$ is the horizon area, $\Phi=-\chi^aA_a$ is horizon electric potential with $A_a$ is the Maxwell field produced by the electric charge $Q$ of the BH and $\kappa$ is the surface gravity. Note that many of these quantities are defined for an asymptotic observer or have a global meaning---this is clear for $M$, $J$ and $Q$; $\Phi$ can be
interpreted as the difference in electrostatic potential between the horizon and infinity, $\Omega$ is the angular velocity of
the horizon as seen from infinity, and $\kappa$ (if
extrapolated from the non-rotating case) is the acceleration of the stationary
observers as they approach the horizon as seen from infinity.

The aim of this letter is to construct a local form of the first law of black hole mechanics. 
For this it will be crucial to describe physics from the viewpoint of our family of observers $\sO$. 

The first situation that we will consider involves the process of absorption  of a test particle by the BH.
More precisely, one throws a test particle of unit mass and charge $q$ from infinity to the horizon. The
geometry is stationary and axisymmetric as well as the electromagnetic field, namely 
${\sL}_{\xi}g_{ab}={\sL}_{\psi}g_{ab}={\sL}_{\xi}A_a={\sL}_{\psi}A_a=0$.
The particle satisfies the Lorentz force equation
\begin{eqnarray}
w^a\nabla_a w_b={q}\,F_{bc}w^c,
\end{eqnarray}
with four-velocity $w^a$. 
The conserved energy of the particle is ${\cal E}\equiv-w^a
\xi_a-qA^a\xi_a$ while the conserved angular momentum is
$L\equiv w^a \psi_a+qA^a\psi_a$. As the particle gets absorbed, the
black hole settles down to a new state with $\delta M={\cal E}$, $\delta J=L$ and $\delta Q=q$.
Equation  (\ref{1st}) then implies
\begin{eqnarray}\label{vani}
\frac{\kappa}{8\pi }\delta A={\cal E}-\Omega L-\Phi q.
\end{eqnarray}
For our observers having four-velocity $u^a$ the local energy of the particle is 
\begin{eqnarray}
{\cal E}_{\ell oc}&=&-w^au_a.\end{eqnarray}
Using (\ref{keyy}), the definitions of $\cal E$, $L$ and $\Phi\equiv -\chi^aA_a$ we find
\ba {\cal E}_{\va \ell oc}=-\frac{w^a\xi_a+\Omega
w^a\psi_a}{\|\chi\|}=\frac{{\cal E}-\Omega
L-q\Phi}{\|\chi\|}.\ea
Finally from (\ref{vani}) 
\ba {\cal E}_{\va loc}
&=&\frac{\overline \kappa}{8\pi}\delta A,
\label{lolo}
\quad{\rm where}\quad
\overline \kappa\equiv\frac{\kappa}{\|\chi\|}.
\ea
From the point of view of our local observers, the horizon has absorbed a particle of energy ${\cal E}_{\ell oc}$. The change in energy of the system $E$ as seen by $\sO$ must be $\delta E={\cal E}_{\ell oc}$.
  All this imply a local version of the first law
\begin{eqnarray}\label{1stlo}
\delta E =\frac{\overline \kappa}{8\pi}\delta A.
\label{firstlaw}
\end{eqnarray}
Direct calculations show that
\footnote{This result follows from the fact that 
$\kappa=\frac{1}{2}({r_+-r_-})/({r_+^2+a^2})$,  
$\chi\cdot \chi= -{\Delta\Sigma_+}/{(r_+^2+a^2)^2}+o(\Delta^{3/2})$, and the proper length to the horizon is
\[\ell=2\sqrt{\frac{\Sigma_+(r-r_+)}{r_+-r_-}}+o(\Delta^{3/2}),\]
 where
$\Delta=(r-r_+)(r-r_-)$ and $\Sigma_+=r_+^2+a^2 \cos^2(\theta)$. }
\be
\overline \kappa\equiv\frac{\kappa}{||\chi||}=\frac{1}{\ell}+o(\ell).
\ee
In other words the local surface gravity measured by the locally non-rotating stationary observers is universal, i.e., independent of the mass $M$, angular momentum $J$ and charge $Q$ of the Kerr-Newman black hole (for a different local definition of surface gravity see \cite{Jacobson:2008cx}). 
Integrating (\ref{1stlo}) we get  the local notion of energy
\be\label{marca}
 E =\frac{A}{8\pi\ell}+o(\ell).
\ee
The idea is to associate the above energy and first law to the horizon itself by taking our $\ell$
as small  as possible without being zero.  An effective quantum gravity formulation where  
thermodynamics makes sense suggest that $\ell$ should be of the order of the Planck scale  \cite{Ghosh:2011fc}
but this is not really essential for the analysis presented here.


A stronger (and local) field theoretic version of the previous arguments goes as follows:
Let the matter falling into the BH be described by a small perturbation of the energy-momentum tensor $\delta T_{ab}$ whose back-reaction to the geometry will be accounted for in the linearized approximation of Einstein's equations around the stationary black hole background. The current $J^a=\delta T^a{}_b\chi^b$ is conserved, $\nabla_aJ^a=0$.
Applying Gauss's law to the spacetime region bounded by the BH horizon $\sH$ and the timelike world-sheet of the observers $\sO$ ($W_{\sO}$) 
we get
\begin{eqnarray}
\int_{\sH} dV dS \ \delta T_{ab}\chi^ak^b=\int_{W_\sO} J_bN^b 
\end{eqnarray}
where $N^a$ is the inward normal of $W_{\sO}$ and $k=\partial_V$ a null normal on $\sH$, with $V$ an affine parameter
along the generators of the horizon. We have also assumed that $\delta T_{ab}$ vanishes in the far past and far future of the considered region.
Using the fact  that $\chi^a=\kappa V k^a$ on $\sH$, the previous identity takes the form
\begin{eqnarray}\label{equis}
\kappa \int_{\sH} dV dS \ V  \delta T_{ab}k^ak^b=\int_{W_\sO} \norma{\chi} \delta T_{ab}u^aN^b ,
\label{equi2}
\end{eqnarray}
Notice that the integral on the right is closely related to the energy-flux associated to the observers, which is equals to $\delta E$. 
Now, the Raychaudhuri equation in the linear approximation is 
\begin{eqnarray}\label{aricha}
\frac{d\theta}{dV}=-8\pi\delta T_{ab}k^ak^b,
\end{eqnarray}
where $\theta$ is the expansion of the null generators $k^a$.
Finally, using the fact that $\| \chi \|$ is constant up to first order on the r.h.s. of (\ref{equis}) we obtain 
\begin{eqnarray}\label{equi}
 \int_{\sH} dV dS \ V  \frac{d\theta}{dV}=-\frac{8\pi {\norma \chi}}{\kappa} \delta E,
\end{eqnarray}
where we have neglected terms of the form $o(\ell) \delta$ which are higher order in our treatment.
By an integration by parts the integral on the left is equal to $-\delta A$.
Finally, using $\overline \kappa\equiv \kappa/{\norma \chi}$ we get the desired result
\begin{eqnarray}
\delta E=\frac{\overline \kappa}{8\pi}\delta A.
\end{eqnarray}
The previous local field theoretical argument will be generalized to include IHs at the end of this letter.


\subsection{On the local temperature at the horizon}

In the usual global framework the surface gravity $\kappa$ is in direct correspondence with the temperature of the Hawking radiation at infinity emitted when quantum effects are taken into account. Does the local surface gravity $\overline \kappa$ have a thermodynamic interpretation if quantum effects are included?
As we argue now, $\overline \kappa$ is in direct correspondence with the temperature of the radiation detected by our family of
local observers.

Suppose that our observers receive a particle with wave four-vector $k_a$. The local frequency $\omega$ they measure is
\be
\omega=k_au^a=\frac{k_a\xi^a+\Omega\,k_a\psi^a}{{\norma \chi}}.
\ee
The geometric optics approximation of the field equations of a charged test field (namely $k^a\nabla_ak_b=qF_{bc}k^c$), 
and the Killing equation and the symmetries of $A_a$, imply that $k_a\xi^a+q A_a\xi^a$ and  $k_a\psi^a+q A_a\psi^a$ are conserved quantities along the particle trajectory. Thus the previous
expression can be written as
\be\label{key}
\omega=k_au^a=\frac{\omega_{\infty}-\Omega m_{\infty}-q\Phi }{{\norma \chi}},
\ee
where $\omega_{\infty}$ and $m_\infty$ are the frequency and angular momentum
of the particles as measured at infinity.

As it is well known Kerr-Newman BHs do not exactly radiate with a Planckian law. According to
the usual results \cite{Hawking:1974sw} the number of radiated particles behaves as
\ba
\langle N_{\infty} \rangle
&=&
 \frac{\Gamma^{\infty}}{\exp\left(\frac{2\pi}{\kappa}[\omega_{\infty}-\Omega m_{\infty}-q \Phi]\right)-1},
\ea
 where $\Gamma^{\infty}$ is the a grey body coefficient
 \footnote{Notice that according to the analysis of \cite{Ghosh:2011fc} one has a definite prediction from LQG concerning the effect of
 quantum geometry on the spectrum of BH radiation at infinity. It is shown that the  first law gets and additional term (a quantum geometry correction) of the form $\mu \delta N$---where 
 $N$ is the number of punctures contributing to the macroscopic geometry of the horizon, and $\mu=-\kappa \sigma(\gamma)/(2\pi)$ with $\sigma(\gamma)$ is a certain function of the Immirzi parameter $\gamma$ \cite{Ghosh:2011fc}.
 Consequently, the BH radiation spectrum would be
 \[ \langle N_{\infty} \rangle
=
 \frac{\Gamma^{\infty}}{\exp\left(\frac{2\pi}{\kappa}(\omega_{\infty}-\Omega_{H} m_{\infty}-q \Phi)+\sigma(\gamma) n\right)-1},\]
 where $n$ is the number of punctures created (distroyed) on the horizon when an excitation of the field of interest is absorbed (emitted) by the BH.  
 The latter quantity would require a more detailed understanding of matter couplings in LQG. A naive interpretation of fermion coupling might suggest that $n=1$ for fermions while
 $n$ could vanish for photons and gravitons.
 }.
The previous expression becomes a standard Planckian spectrum when expressed in terms of the
local frequency measured next to the horizon, namely
\ba
\langle N_{\infty} \rangle
&=&
  \frac{\Gamma^{\infty}}{e^{\frac{2\pi {\norma \chi}}{\kappa}\omega}-1},
\ea
 where we have replaced (\ref{key}).
 This transmission coefficient is assumed to behave properly (slowly varying in the suitable ranges) and does not interfere with the thermal nature of the outgoing radiation. The previous equation and usual arguments that can be made by looking at the details of Hawking's calculation imply that local observers at the horizon
 detect the following spectrum
 \be
\langle N \rangle
=
 \frac{\Gamma}{e^{\frac{2\pi}{\overline \kappa}\omega}-1},
\ee
 where we have used the definition of the local surface gravity (\ref{lolo}).
 In other words local observers detect a genuine Planckian spectrum at temperature $\overline \kappa/(2\pi)$!
 


\subsection{Isolated horizons}

Here we prove the validity of the local form of the first law (\ref{1stlo}) in the more general framework of IHs. Moreover, the first law so derived is dynamical in character, i.e., changes in the area and energy of the system can really be seen as the consequence of the absortion of matter fields by the horizon along its history. 
Isolated horizons are equipped with an equivalence class of null normal $[\chi]$ where equivalence is defined up to constant scalings.
The generators $\chi_a$  are geodesic and define a notion of isolated horizon surface gravity $\kappa_{\va IH}$ through the equation
 $\nabla_{\chi} \chi_a=\kappa_{\va IH} \chi_a$. It is clear that $\kappa_{\va IH}$ is not defined in $[\chi]$ because it gets rescaled when $\chi$ is rescaled. 
The near horizon geometry is described (in terms of Bondi like coordinates) by the metric \cite{prl}
\ba\n
ds^2=2 dv_{(a}dr_{b)}&-&2(r-r_0) [2dv_{(a}\omega_{b)}-\kappa_{\va IH} dv_adv_b ] \\ &+& q_{ab}+o[(r-r_0)^2],
\ea
where $\chi=\partial_v$ is the extension to the vicinity of $\sH$ of the null generators of the IH through the
flow of a natural null vector $n^a=\partial^a_r$ (i.e. $\sL_n(\chi)=0$), $q_{ab}n^a=q_{ab}\chi^a=0$, and $\omega_a$ is a one 
form intrinsic to IHs with the important property that $\omega(\chi)|_{\sH}=\kappa_{\va IH}$. 
Also one has  \cite{gr-qc/0101008}
\be\sL_{\chi}g_{ab}|_{\sH}=0.\ee
Thus $\chi$ can be used to define the observers $\sO$ as in (\ref{keyy}).  The proper distance $\ell$ to the horizon from a point with coordinate $r$
along a curve normal both to $\chi$ and $q_{ab}$---with tangent vector $N^a=\partial^a_r+({2\kappa_{\va IH} (r-r_0)})^{-1} \partial^a_v
$---is given by 
$\ell=\sqrt{{2(r-r_0)}/\kappa_{\va IH}}$, while $\chi\cdot\chi=2\kappa_{\va IH} (r-r_0)$.  Therefore, 
\be
\overline \kappa=\frac{\kappa_{\va IH}}{||\chi||}=\frac{1}{\ell}.
\ee
Notice that $\overline \kappa$ is well defined in $[\chi]$ in contrast with $\kappa_{\va IH}$. 
As the form of the perturbed Raychaudhuri equation (\ref{aricha}) is the same for the generators of IH (as their expansion, shear, and twist vanish by definition),  the same arguments given below eq. (\ref{marca})
yield the local first law
\begin{eqnarray}\label{1stih}
{\delta E= \frac{\overline \kappa}{8\pi}\delta A},
\end{eqnarray} where the energy notion $E=\frac{A}{8\pi \ell}$,  and we have used that $\ell^2<<A$, provide the right local framework  for the
statistical mechanics study of quantum IHs.

Summarising: even though we have first justified the local first law (\ref{1stih}) starting from the
analysis of the first law for stationary spacetimes and its translation in terms of the 
local observers $\sO$, the final analysis for IHs implies that the result can be recovered entirely from local considerations that know nothing about the global structure.
In this letter we are proposing to use this remarkable fact in order to reverse perspective, and thus take the
local definition of IHs with its null normals $[\chi]$, the local first law (\ref{1stih}), and its associated energy
as the fundamental structure behind BH thermodynamics. 

Notice also that the local first law and the universality of $\overline\kappa$ implies the Gibbs relation
$E=T  S$
where $T={\ell_p^2\overline \kappa} / ({2\pi})$, and $S={A}/{4\ell_p^2}$.
This simple property of usual thermodynamic systems is not realized by the quantities taking part in the standard first law (\ref{1st}).
{This is an extra bonus of our local description}.

We are grateful for exchanges and remarks with F. Barbero,  S. Dain, C. R\"{o}ken, and E. Wilson-Ewing.
We thank {\em l'Institut Universitaire de France} for support. EF was supported by CONICYT (Chile) grant D-21080187 and the MECE (Chile) program.


\begin{thebibliography}{1}

%

\bibitem{Hawking:1974sw}
  S.~W.~Hawking,
  Commun.\ Math.\ Phys.\  {\bf 43} (1975) 199.

\bibitem{strings}
  A.~Strominger, C.~Vafa,
  Phys.\ Lett.\  {\bf B379 } (1996)  99-104.
(for a review see) A.~W.~Peet,
  [hep-th/0008241].

  \bibitem{Corichi:2009wn}
 C.~Rovelli,
  Phys.\ Rev.\ Lett.\  {\bf 77} (1996) 3288.
  A.~Ashtekar, J.~Baez, A.~Corichi and K.~Krasnov,
  Phys.\ Rev.\ Lett.\  {\bf 80} (1998) 904.
R. K. Kaul and P. Majumdar, Phys.\ Lett.\ {\bf B439} (1998) 267.
    A.~Corichi,
  arXiv:0901.1302 [gr-qc].
M. Domagala and J. Lewandowski, Class.\ Quant.\ Grav. {\bf 21} (2004) 5233.
K. A. Meissner, Class.\ Quant.\ Grav.\ {\bf 21} (2004) 5245.
A. Ghosh and P. Mitra, Phys.\ Lett.\ {\bf B616} (2005) 114.
  J.~Engle, A.~Perez and K.~Noui,
  Phys.\ Rev.\ Lett.\  {\bf 105} (2010) 031302.
   R.~Basu, R.~K.~Kaul and P.~Majumdar,
  Phys.\ Rev.\  D {\bf 82} (2010) 024007.

\bibitem{Ashtekar:2008jd}
  A.~Ashtekar, V.~Taveras and M.~Varadarajan,
  Phys.\ Rev.\ Lett.\  {\bf 100}, 211302 (2008).
  A.~Ashtekar, F.~Pretorius and F.~Ramazanoglu,
  Phys.\ Rev.\ Lett.\  {\bf 106}, 161303 (2011).

\bibitem{Ashtekar:2004cn}
  A.~Ashtekar and B.~Krishnan,
  Liv. Rev.\ Rel.\  {\bf 7}, 10 (2004).

\bibitem{prl}  A.~Ashtekar, C.~Beetle, O.~Dreyer, S.~Fairhurst, B.~Krishnan, J.~Lewandowski, J.~Wisniewski,
  Phys.\ Rev.\ Lett.\  {\bf 85 } (2000)  3564-3567.



\bibitem{aa}
  A.~Ashtekar, S.~Fairhurst and B.~Krishnan,
  Phys.\ Rev.\  D {\bf 62} (2000) 104025.
   A.~Ashtekar, C.~Beetle, J.~Lewandowski,
  Phys.\ Rev.\  {\bf D64 } (2001)  044016.
  



\bibitem{Krasnov:1997yt}
  K.~V.~Krasnov,
  Class.\ Quant.\ Grav.\  {\bf 16} (1999) 563.  J.~F.~B.~G., E.~J.~S.~Villasenor,
  Class.\ Quant.\ Grav.\  {\bf 28 } (2011)  215014.


\bibitem{Wald:1984rg}
  R.~M.~Wald,
  ``General Relativity,''
  Chicago, Usa: Univ. Pr. ( 1984) 491p.
  
  \bibitem{Thorne:1986iy}
  K.~S.~Thorne, (Ed.), R.~H.~Price, (Ed.), D.~A.~Macdonald, (Ed.),
  ``Black Holes: The Membrane Paradigm,''
  NEW HAVEN, USA: YALE UNIV. PR. (1986) 367p.
  
\bibitem{Ghosh:2011fc} 
  A.~Ghosh and A.~Perez,
  Phys.\ Rev.\ Lett.\  {\bf 107}, 241301 (2011).


\bibitem{Jacobson:2008cx}
  T.~Jacobson, R.~Parentani,
  Class.\ Quant.\ Grav.\  {\bf 25 } (2008)  195009.

\bibitem{gr-qc/0101008} 
  J.~Lewandowski and T.~Pawlowski,
  Int.\ J.\ Mod.\ Phys.\ D\ {\bf 11}, 739  (2002)
  [gr-qc/0101008].
  
  
\end{thebibliography}
\end{document}